\documentclass{webofc}

\usepackage[varg]{txfonts}   
\usepackage{hyperref}
\usepackage{url}
\usepackage{todonotes}
\usepackage{regexpatch}
\usepackage{xargs}
\usepackage{graphicx}
\usepackage{subcaption}
\usepackage{booktabs}
\usepackage[export]{adjustbox}
\usepackage{csquotes}
\hypersetup{colorlinks=true,citecolor=blue,urlcolor=blue,linkcolor=blue}
\makeatletter
\xpatchcmd{\@todo}{\setkeys{todonotes}{#1}}{\setkeys{todonotes}{inline,#1}}{}{}

\begin{document}
\title{Surrogate Modeling for Scalable Evaluation of Distributed Computing Systems for HEP Applications}

\author{\firstname{Larissa} \lastname{Schmid}\inst{1}\fnsep\thanks{\email{larissa.schmid@kit.edu}} \and
        \firstname{Maximilian} \lastname{Horzela}\inst{2}\fnsep\thanks{\email{maximilian.horzela@cern.ch}} \and
        \firstname{Valerii} \lastname{Zhyla}\inst{1}\fnsep \and
        \firstname{Manuel} \lastname{Giffels}\inst{3}\fnsep \and
        \firstname{Günter} \lastname{Quast}\inst{3}\fnsep \and
        \firstname{Anne} \lastname{Koziolek}\inst{1}\fnsep
}

\institute{KASTEL - Institute of Information Security and Dependability, Karlsruhe Institute of Technology
\and
           II.~Institute of Physics, Georg-August Universität Göttingen
\and
           Institute of Experimental Particle Physics (ETP), Karlsruhe Institute of Technology
          }

\abstract{
The Worldwide LHC Computing Grid (WLCG) provides the robust computing infrastructure essential for the LHC experiments by integrating global computing resources into a cohesive entity. Simulations of different compute models present a feasible approach for evaluating future adaptations that are able to cope with future increased demands. However, running these simulations incurs a trade-off between accuracy and scalability. For example, while the simulator DCSim can provide accurate results, it falls short on scaling with the size of the simulated platform. Using Generative Machine Learning as a surrogate presents a candidate for overcoming this challenge.

In this work, we evaluate the usage of three different Machine Learning models for the simulation of distributed computing systems and assess their ability to generalize to unseen situations. We show that those models can predict central observables derived from execution traces of compute jobs with approximate accuracy but with orders of magnitude faster execution times. Furthermore, we identify potentials for improving the predictions towards better accuracy and generalizability.
}

\maketitle

\section{Introduction}
\label{sec:intro}High energy physics at the LHC relies on the global-scale federated computing infrastructure provided by the Worldwide LHC Computing Grid (WLCG)~\cite{wlcg-web} for processing and storing the sheer amounts of scientific data gathered by the LHC experiments.
To achieve the high-throughput demands with limited financial resources, good performance of the computing system has to be constantly ensured.
With increasing demands expected from future LHC operations and technical and strategic changes in the federation, the best design for the WLCG infrastructure still needs to be identified.

Due to the size and complexity of such infrastructures, it is not feasible to build alternative infrastructures for mere testing.
In addition, due to constant uptime requirements, they cannot be reserved for test and evaluation purposes.
Instead, accurate simulation of workflow executions on such infrastructures and subsequent analysis of the simulated results can be performed without disturbing the operation of the real reference infrastructure, and a variety of alternative designs can be tested with just a fraction of the original investment.

However, there is a general trade-off between accuracy and the computational complexity, and therefore execution time, that comes with designing practical simulation models as simulations with higher accuracy generally implement more fine-granular models with greater detail.
This typically leads to superlinear scaling in the execution time of the respective simulation tools with respect to the size of the simulated infrastructure.
As a result, it quickly becomes unfeasible to simulate workflow executions on large infrastructures.

To circumnavigate the unfavorable scaling of such simulation tools, ML surrogate models, trained on verifiable accurate simulator outputs or data from real-world systems, that predict observables from job execution traces in constant time present a natural solution approach.
\section{Data Generator DCSim}
\label{sec:background}

To simulate the execution of HEP workflows on parallel and distributed computing (PDC) infrastructure, we utilize the DCSim tool~\cite{Horzela2023Proceeding,Horzela2023_1000165566} as a reference simulator\footnote{Note the exact version of the simulator published here: \url{https://doi.org/10.5281/ZENODO.8300961}}. 
It is implemented using the SimGrid~\cite{SimGrid} and WRENCH~\cite{wrench} simulation frameworks.
SimGrid and WRENCH have been chosen since they are general-purpose, enable more accurate simulation of PDC systems than other versatile frameworks~\cite{TOMACS} while keeping the level of computational complexity at manageable levels, have been carefully validated~\cite{
simutool_09, nstools_07, simgrid_storage,
7384330,
Cornebize-cluster19}, and are utilized by a large and active community. 

DCSim takes three main essential user inputs indicated as \enquote{configurations} in \autoref{fig:approach-overview}.
First, a platform description following the SimGrid standard defines the network of computing resources with all its (technical) characteristics, for instance, the network of links and interconnected computers, their routing, as well as their respective network bandwidths, latencies, numbers of CPU cores, CPU speeds, disk bandwidths, storage volumes, and other parameters.
Second, datasets and replicas present at the simulation start, their composition of files, sizes, and locations are specified in a dedicated input file.
Third, the workloads to be executed on the specified platform while possibly processing files from datasets are defined in another dedicated steering file.
The workloads are characterized by their submission time, the job collection they are composed of, their respective resource demands, and the amount of computational work that must be executed for their completion.
Additionally, parameters can be given to DCSim to steer individual components of the simulation models integrated into DCSim or to adjust their calibration. 
In this work, we use the default calibration derived for \cite{Horzela2023Proceeding}.
Using these inputs, DCSim simulates the scheduling of the jobs, their execution and data processing, and returns a list of observables for each simulated job.
Per default, observables are the start and end times of the jobs, the total times spent with I/O and compute operations, and the processed amounts of data.

It has been shown that after calibration, DCSim can achieve valid descriptions of real-world computing workflow executions and systems~\cite{Horzela2023Proceeding,Horzela2023_1000165566}.
However, the execution time of DCSim scales superlinearly with the size of the simulated computing system, which limits its feasibility and necessitates workarounds in studies concerning computing systems at a global scale, like the WLCG.
We therefore utilize DCSim as a baseline for testing ML surrogates in simulation.

\section{Machine Learning Approach}
\label{sec:approach}

\begin{figure}
\begin{minipage}[b]{0.54\linewidth}
    \centering
    \includegraphics[width=\linewidth]{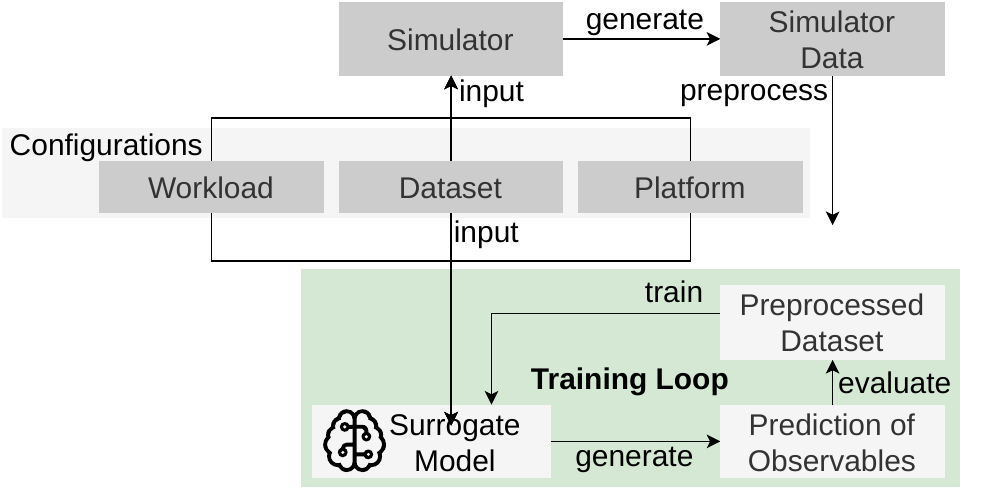}
    \caption{Approach overview.}
    \label{fig:approach-overview}
\end{minipage}
\hfill
\begin{minipage}[b]{0.44\linewidth}
\centering
    \includegraphics[width=\linewidth]{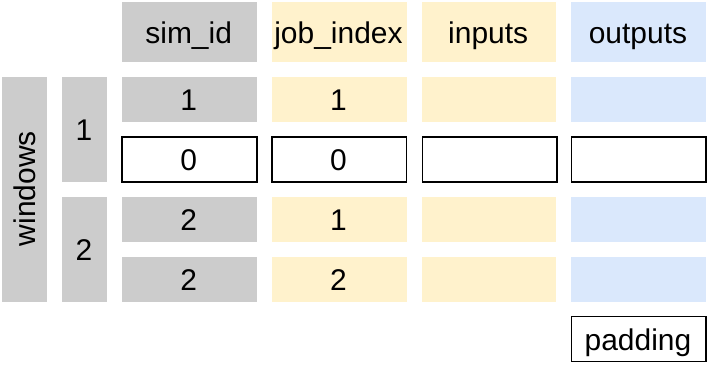}
    \caption{Layout of training dataset.}
    \label{fig:approach-data}
\end{minipage}
\end{figure}

\autoref{fig:approach-overview} shows an overview of our general approach. 
First, we conduct simulations using DCSim and preprocess the resulting data (Section~\ref{sec:approach:preprocessing}). 
We then train our model using the preprocessed simulator data (Section~\ref{sec:approach:training}). The model receives the configuration parameters of the simulation, i.e., the execution platform, characterization of datasets used, and workload configurations, as input and predicts the observables of each job in the workload. 
In total, we train three different model architectures (Section~\ref{sec:approach:architectures}). 
%

\subsection{Data Preprocessing} \label{sec:approach:preprocessing}
To create a sequential dataset that can be used to train our model, we need to preprocess the DCSim output data. 
First, we standardize the data to ensure that all features contribute equally to the predictions of the models, preventing features on larger scales from dominating the learning process. 
For each feature, we separately subtract the mean and divide the result by the standard deviation, thus adjusting the mean to zero and the standard deviation to one. 
%
%
We apply an inverse scaling transformation on the models' outputs to retrieve the output data at their original scale. 

Next, since the number of jobs varies in each simulation and some models require fixed-length input sequences, we divide the model input vectors into windows of fixed size. 
If the last window of a simulation contains fewer jobs than the fixed size, we pad it with zeros to maintain a consistent window size. 
To be able to reconstruct the original data sequence as predicted by the simulations, we add the \texttt{simulation\_ID} (not used by the model) and \texttt{job\_index} (used for model training) as additional features to the entries in each input window.
We provide an example illustration of a windowed dataset in \autoref{fig:approach-data}.

\subsection{Model Training} \label{sec:approach:training}

We split the data into training and evaluation datasets to train the model, using a 70:30 ratio per simulation of the same length. 
We employ sequence-to-sequence prediction, predicting exactly one output per input row. 
%
%
%
%
The mean squared error (MSE) loss function is used to penalize larger errors more heavily than smaller ones, emphasizing the importance of minimizing significant discrepancies between predicted and actual values.

We adjust hyperparameters, including the hidden size, which defines the dimensionality of each hidden layer, the window size specifying the number of consecutive data points considered in each input window, and the window overlap, determining how much consecutive windows overlap. 
Window overlap between windows from the same simulation can be used to capture the continuity and dependencies between jobs within the same simulation, enhancing the model's ability to learn from sequential patterns. 
In addition, we tune the batch size, which indicates the number of input rows processed before updating the internal parameters of the model, and the number of hidden layers in the model.

We optimize these hyperparameters manually by evaluating kernel density estimation (KDE) and accuracy plots. 
We begin by training ten models with randomly chosen hyperparameter configurations and select the three best-performing models. 
For each of these models, we adjust one free hyperparameter at a time, train the model, and evaluate its performance. 
Once the optimal value for a specific parameter is identified, we fix it and continue with the following free parameter while retaining the best-performing model from the previous iteration. 
This process continues until all hyperparameters have been optimized. 
Finally, we compare the three resulting models and select the one that demonstrates the best overall performance. 
The tuning order we follow is hidden size, window size, window overlap, number of layers, and batch size.

\subsection{Model Architectures} \label{sec:approach:architectures}

We explore three different model architectures for our surrogate model: BiGRU, BiLSTM, and Transformer models.

The BiGRU architecture consists of a linear input layer, multiple bidirectional Gated Recurrent Unit (BiGRU) layers, and a single linear output layer~\cite{salehinejad_recent_2018, yuanshuai_duan_improved_2023}. 
BiGRU processes data in both forward and backward directions, making it effective in scenarios where both past and future contexts help understand the sequence. 
Although future jobs cannot influence already completed jobs, the training phase can benefit from data on future jobs by helping the model identify patterns or dependencies that might not be evident when considering only past and current jobs. 
Moreover, BiGRU can capture long-term dependencies, which is valuable for predicting simulation results, as it allows the model to account for the influence of events over extended periods. 
This capability can lead to more accurate predictions since jobs submitted earlier and still running can influence subsequent jobs.

The BiLSTM architecture has a structure similar to that of BiGRU but uses bidirectional Long Short-Term Memory (BiLSTM) layers instead of BiGRU layers~\cite{khaled_a_althelaya_evaluation_2018, salehinejad_recent_2018}. 
As BiGRU, BiLSTM can capture long-term dependencies and processes data in both forward and backward directions.

The transformer-based architecture we use~\cite{wu_deep_2020} is adapted for numerical predictions. 
Apart from technical scalability and performance benefits, thanks to the self-attention mechanism, transformers can handle long-range dependencies and are suited better for handling local and global contexts simultaneously.
In this study, we focus on encoder-only transformers, with the number of attention heads as an additional hyperparameter.

\section{Evaluation of Surrogate Models}
\label{sec:eval}

We evaluate our approach by varying the complexity of the data used for model assessment and comparing the predicted results with the ones obtained with DCSim. 
In the first evaluation scenario, referred to as homogeneous jobs (Section~\ref{sec:eval:homogeneous}), we use jobs generated from a single workload description executed on a fixed platform with minimal complexity. 
This controlled setup allows us to establish a baseline for the model's performance under simple conditions. 
The second scenario covers heterogeneous jobs (Section~\ref{sec:eval:heterogeneous}), introducing increased complexity by predicting results for jobs derived from multiple workload distributions on a more complex platform. 
This setting provides insight into the model's capabilities to handle diverse workloads. 

We show the evaluation results for the observables \textit{compute\_time} and \textit{input\_files\_transfer\_time} for both scenarios. The results for other observables can be found in~\cite{Zhyla2024}. 

We evaluate the models' predictions using the coefficient of determination (R-squared) and Kernel Density Estimate (KDE) plots. 
R-squared represents the proportion of variance in the target variable explained by the model. Its values range from zero to one, with one indicating a perfect fit. 
KDE plots provide a visual assessment of the distribution of the respective observable, revealing patterns or systematic bias regarding under- or overprediction that R-squared cannot capture. 

\subsection{Homogeneous Jobs} \label{sec:eval:homogeneous}

In the homogeneous jobs scenario, we use 10,000 simulations for training, organized into 10 batches of 1,000 simulations each. 
These simulations include 1, 10, 20, 50, 100, 250, 500, 1,000, 1,500, and 2,000 jobs per simulation. 
To assess extrapolation capabilities, we use an additional dataset of 10 simulations with 10,000 jobs each that are not used in training. 
The resource demands of all jobs remain identical throughout the training and evaluation phases.
As input features, we use \textit{index}, \textit{flops}, \textit{input\_files\_size}, and \textit{output\_files\_size}. 
The platform configuration contains three worker nodes, two with 24 CPU cores each, one with 12 CPU cores, and one scheduler node. 
All nodes are connected following a star topology. 
The datasets processed by the jobs are stored on a separate storage server. 

\begin{figure}%
    \begin{subfigure}{.49\linewidth}%
    \centering%
    \includegraphics[width=\linewidth]{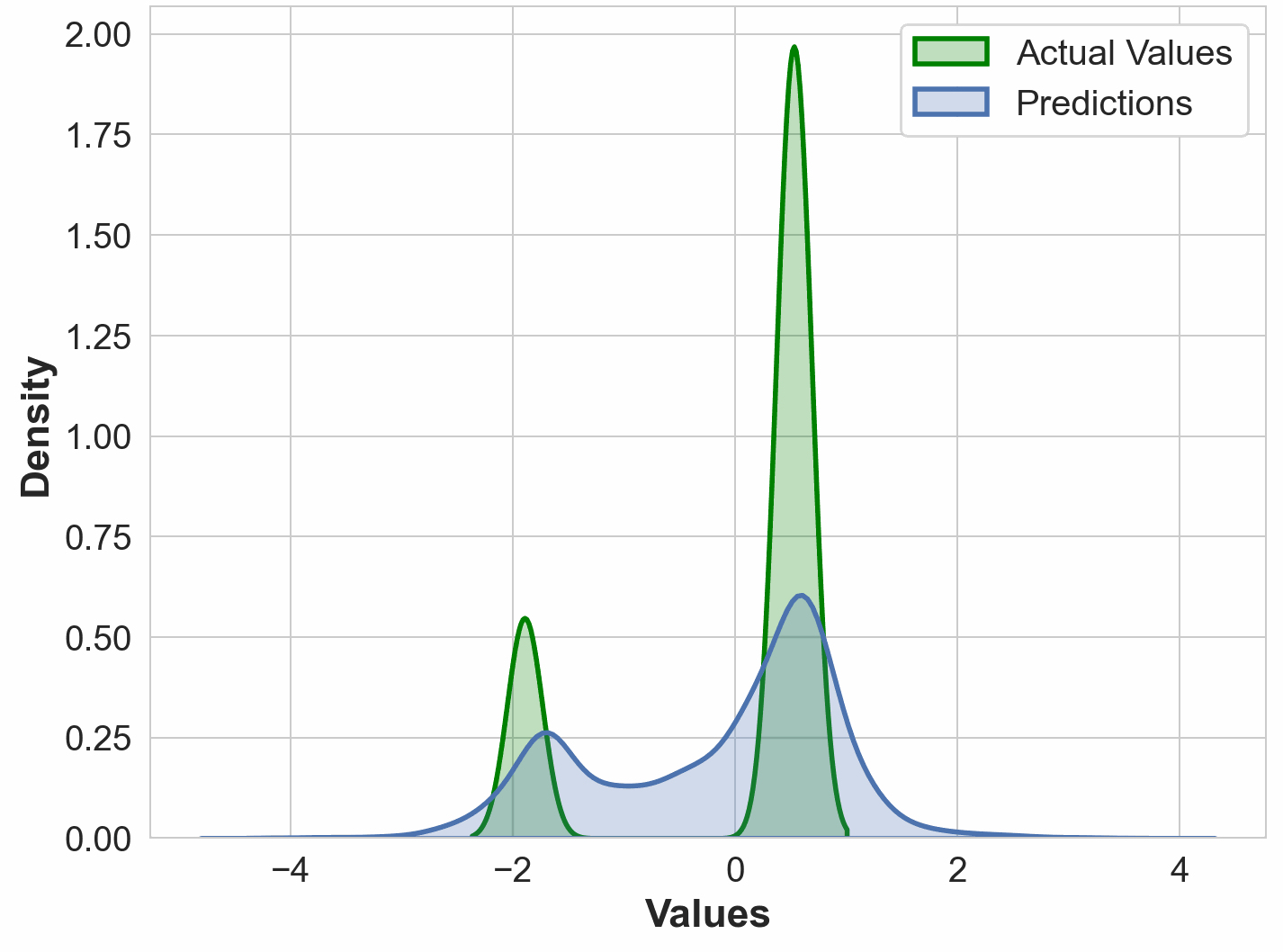}%
    \vspace{-0.5em}%
    \caption{BiGRU model, \textit{compute\_time}.}%
    \label{fig:eval-1-gru-extrapolation-compute}%
    \end{subfigure}%
    \hfill%
    \begin{subfigure}{0.49\linewidth}%
    \centering%
    \includegraphics[width=\linewidth]{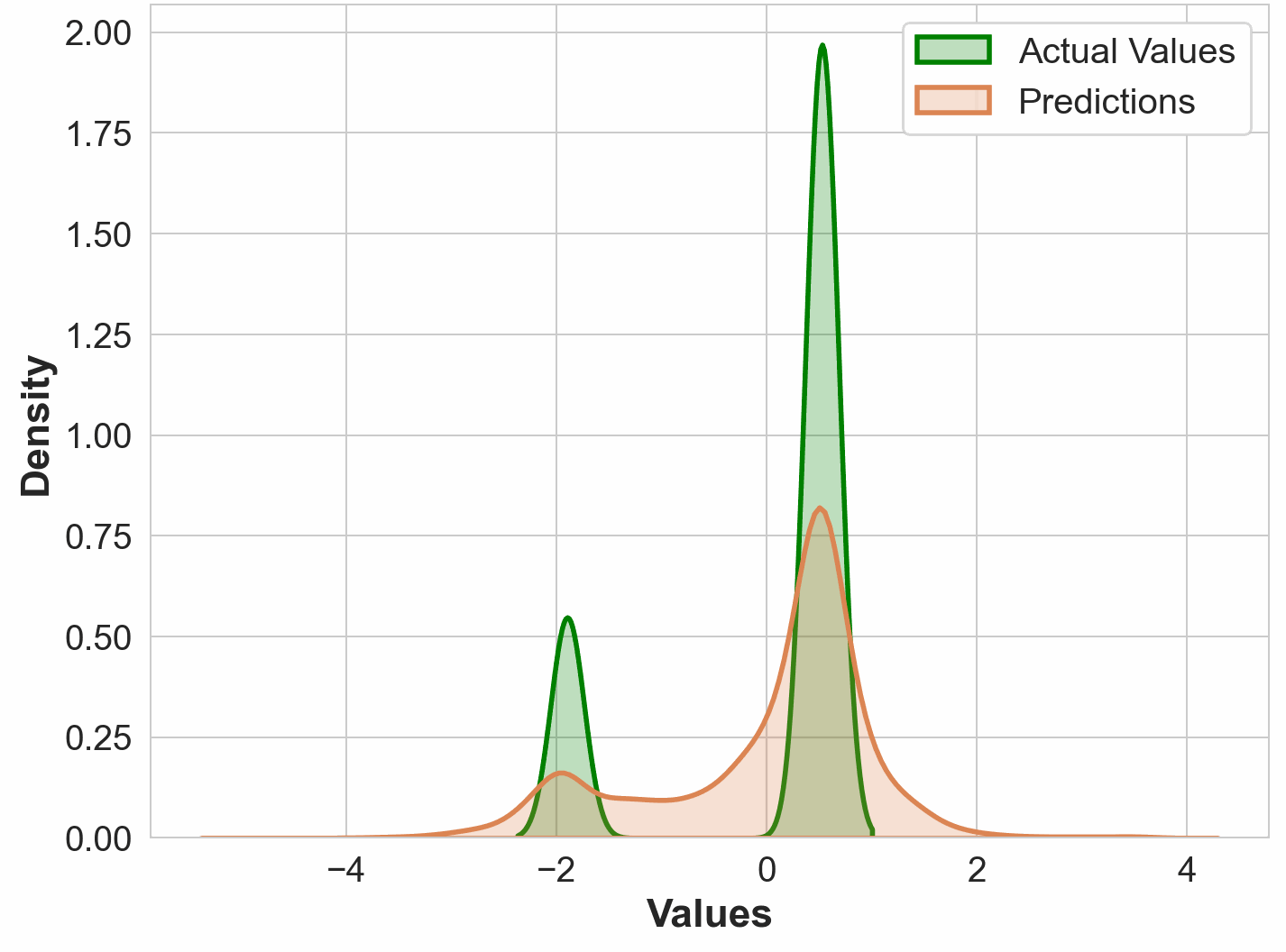}%
    \caption{BiLSTM model, \textit{compute\_time}.}%
    \label{fig:eval-1-lstm-extrapolation-compute}%
    \end{subfigure}%
    \hfill%
    \vspace{-0.5em}
    \begin{subfigure}{.49\linewidth}%
    \centering%
    \includegraphics[width=\linewidth]{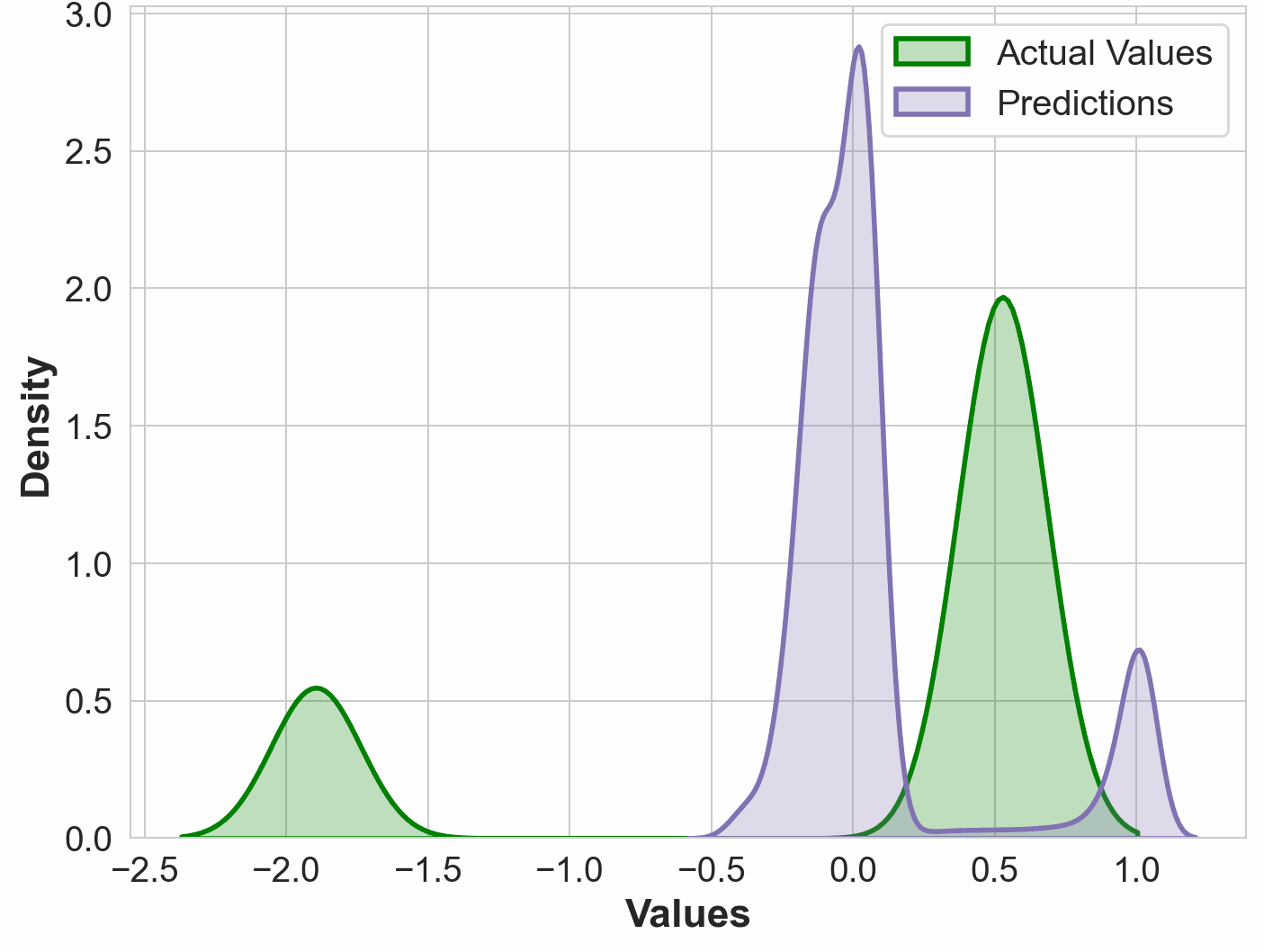}%
    \vspace{-0.5em}%
    \caption{Transformer model, \textit{compute\_time}.}%
    \label{fig:eval-1-transformer-extrapolation-compute}%
    \end{subfigure}%
    \hfill%
    \begin{subfigure}{.49\linewidth}%
    \centering%
    \includegraphics[width=\linewidth]{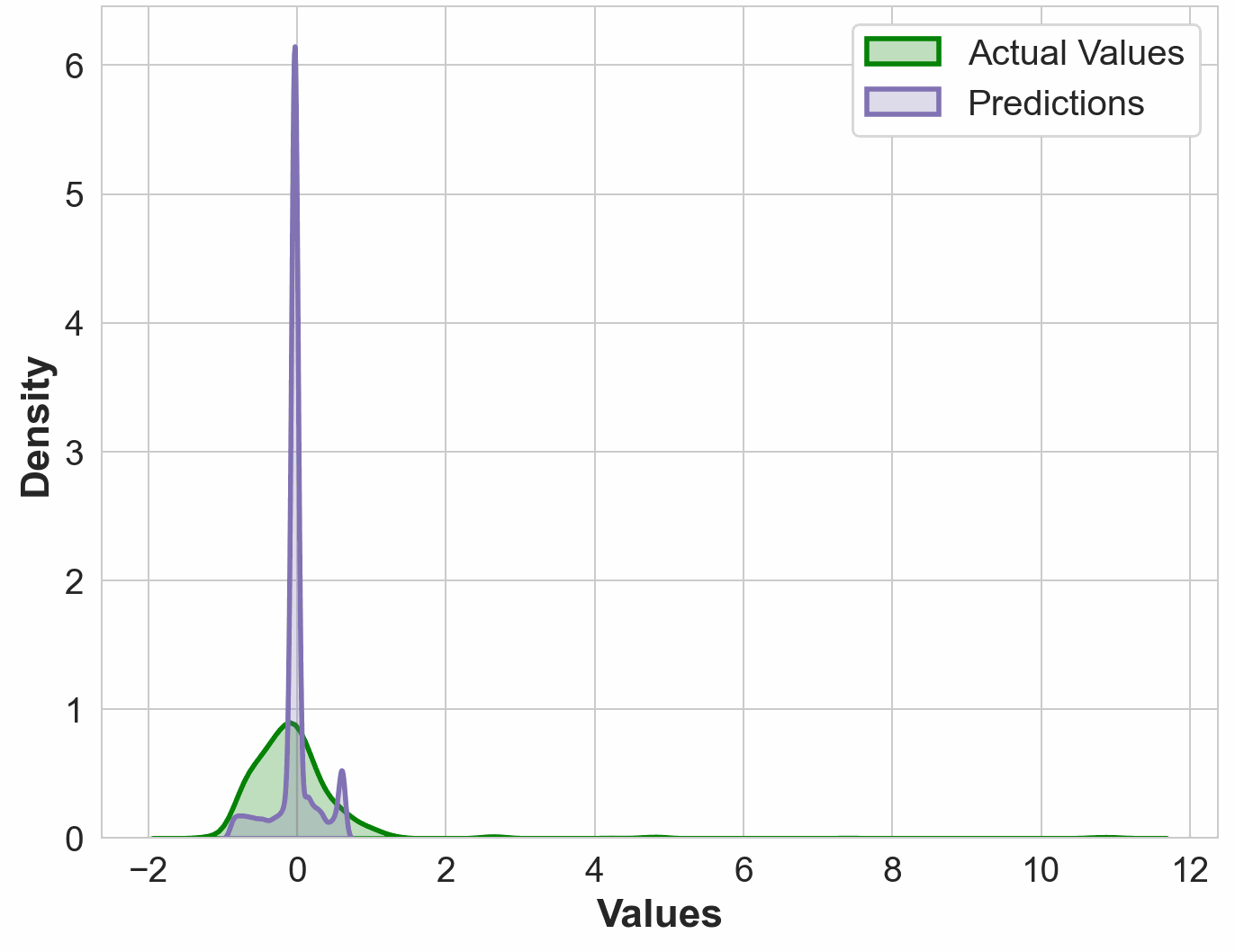}%
    \vspace{-0.5em}%
    \caption{Transformer model, \textit{input\_files\_transfer\_time}.}%
    \label{fig:eval-1-transformer-extrapolation-input_files_transfer_time}%
    \end{subfigure}%
    \caption{Predictions for extrapolation of trained models in the homogenous jobs setup. The \textit{input\_files\_transfer\_time} is similar for the other models.}%
    \label{fig:eval-1}%
\end{figure}

We show a comparison of the prediction results with the simulation for the extrapolation task in \autoref{tab:eval} and \autoref{fig:eval-1}. 
%
%
All models exhibit a negative R-squared value for \textit{compute\_time}, indicating a worse fit than a naive baseline. 
The KDE plots show that the target distribution exhibits two prominent spikes for the \textit{compute\_time}. While the BiGRU and BiLSTM models successfully capture the general structure of the distribution, they fail to emphasize these two spikes sufficiently. 
The transformer model, however, predicts mainly the mean values for \textit{compute time}, failing to capture the actual distribution. 
While the R-squared value for \textit{input\_files\_transfer\_time} is positive, it is close to zero for all models, indicating only a slightly better fit than the naive baseline. 
The KDE plots show that the models capture some distribution features again but fail to emphasize the actual shape. 
In summary, all models failed to predict the observables beyond the first statistical moment, potentially due to the lack of input features providing additional context constraining the job execution, particularly information about the platform architecture.
Although the size of the modelled infrastructure is rather minimal in this example, the surrogate ansatz already leads to improvements in the execution time of minimum two orders of magnitude on our test systems ($\mathcal{O}(10\,\mathrm{s}) \to \mathcal{O}(100\,\mathrm{ms})$).

\subsection{Heterogeneous Jobs} \label{sec:eval:heterogeneous}

Our second evaluation scenario introduces jobs from multiple distributions and a more complex platform configuration.
We train the models and evaluate their extrapolation capabilities using the same strategy as for the homogeneous jobs.
The workload comprises five distinct job classes, each with its own distribution for all resource demands.
Each job's \textit{submission\_time} differs in this scenario and is therefore added as an input feature.
The platform configuration describes an interconnected network of two data centers. 
The first consists of a local network of ten identical worker nodes and a storage node. 
The second has a single but more powerful worker node. 
All datasets are stored in the first data center. 
The first data center features 420 CPU cores, while the second has 200 CPU cores.

\begin{figure}%
    \begin{subfigure}{0.49\linewidth}%
    \centering%
    \includegraphics[width=\linewidth]{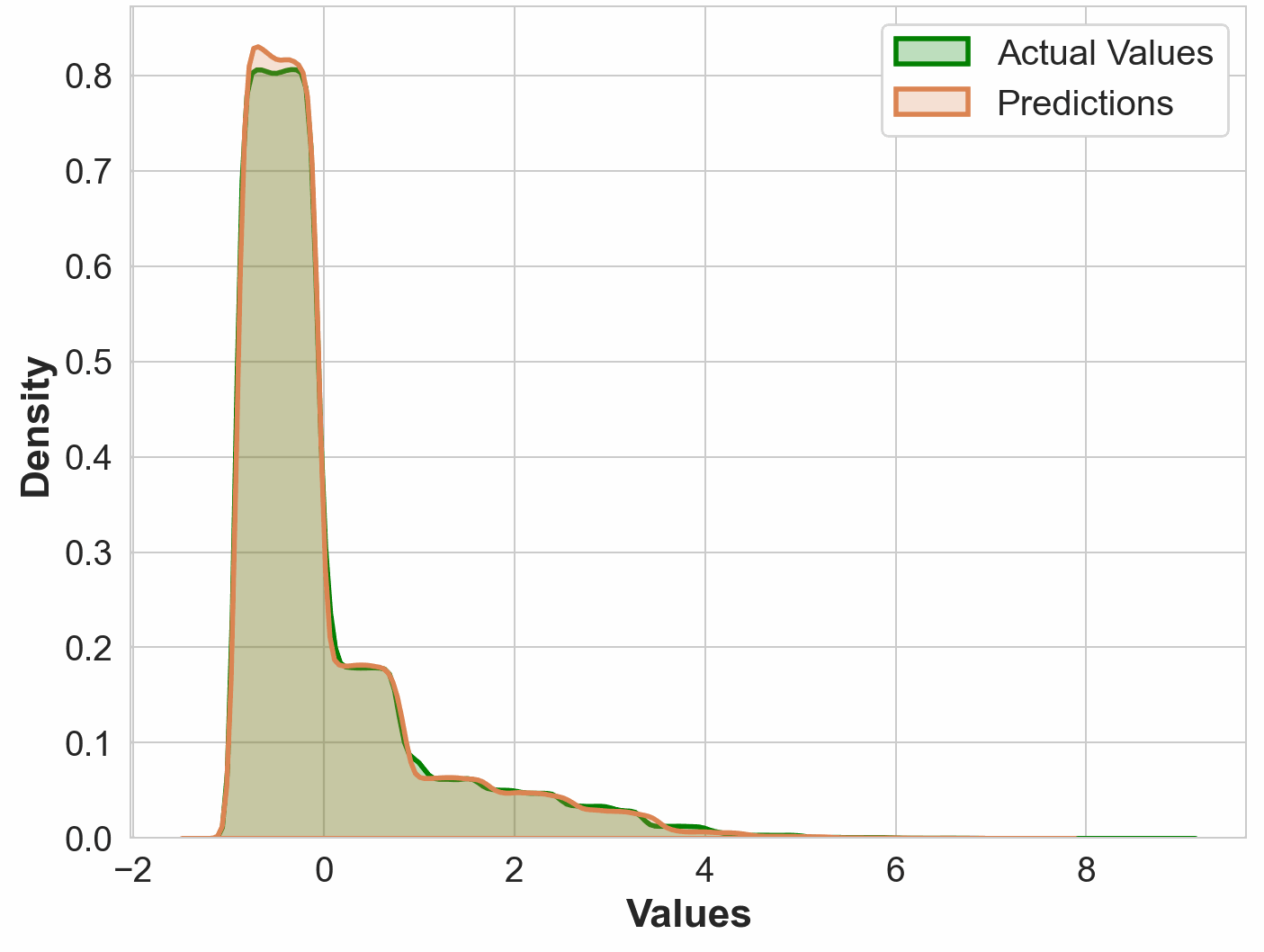}%
    \vspace{-0.5em}%
    \caption{BiLSTM model, \textit{compute\_time}}%
    \label{fig:eval-3-LSTM-extrapolation-compute}%
    \end{subfigure}%
    \hfill%
    \begin{subfigure}{0.49\linewidth}%
    \centering%
    \includegraphics[width=\linewidth]{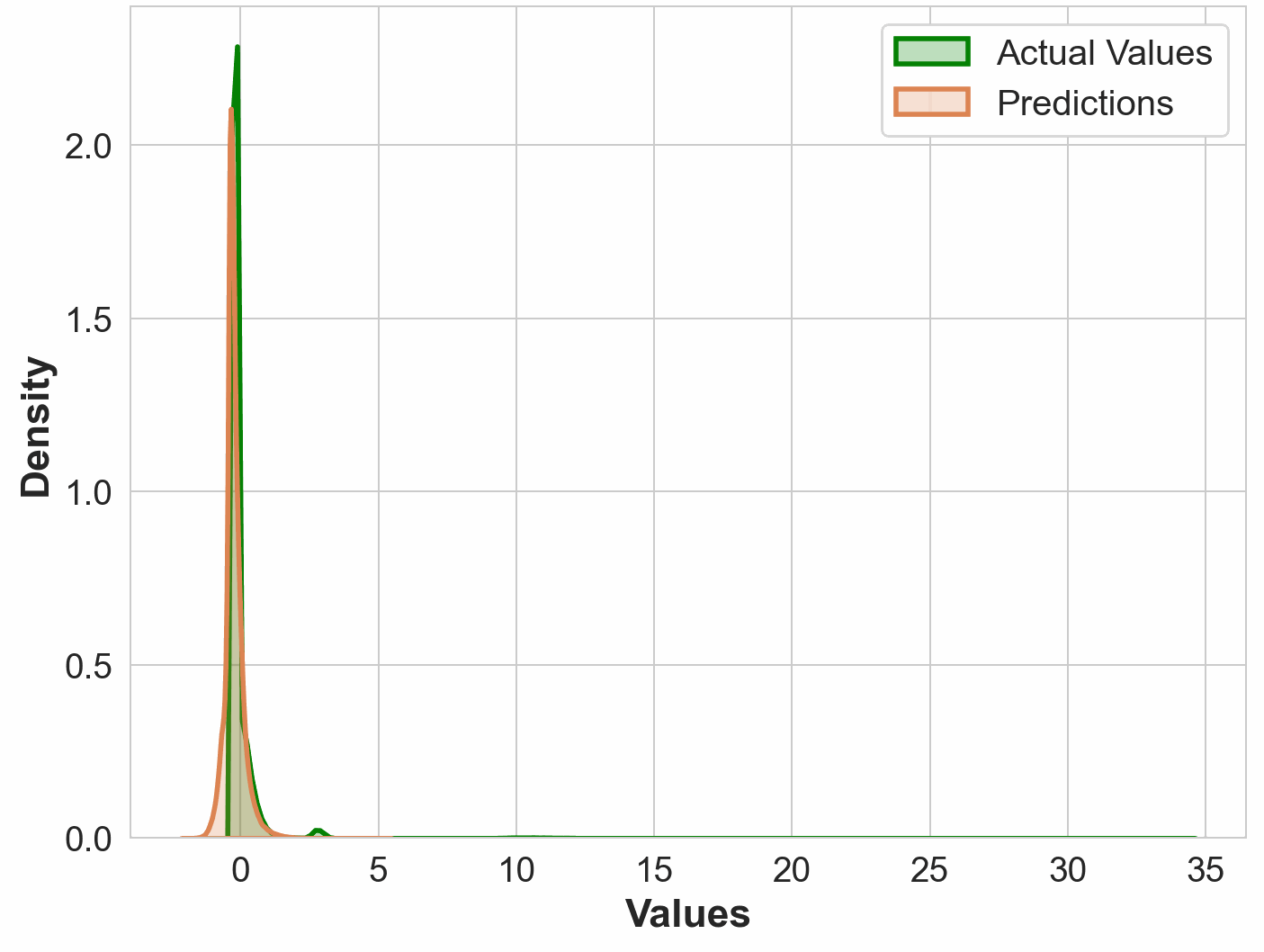}%
    \vspace{-0.5em}
    \caption{BiLSTM model, \textit{input\_files\_transfer\_time}}%
    \label{fig:eval-3-LSTM-extrapolation-input}%
    \end{subfigure}%
    \label{fig:eval-3-lstm}%
    \hfill
    \begin{subfigure}{0.49\linewidth}%
    \centering%
    \includegraphics[width=\linewidth]{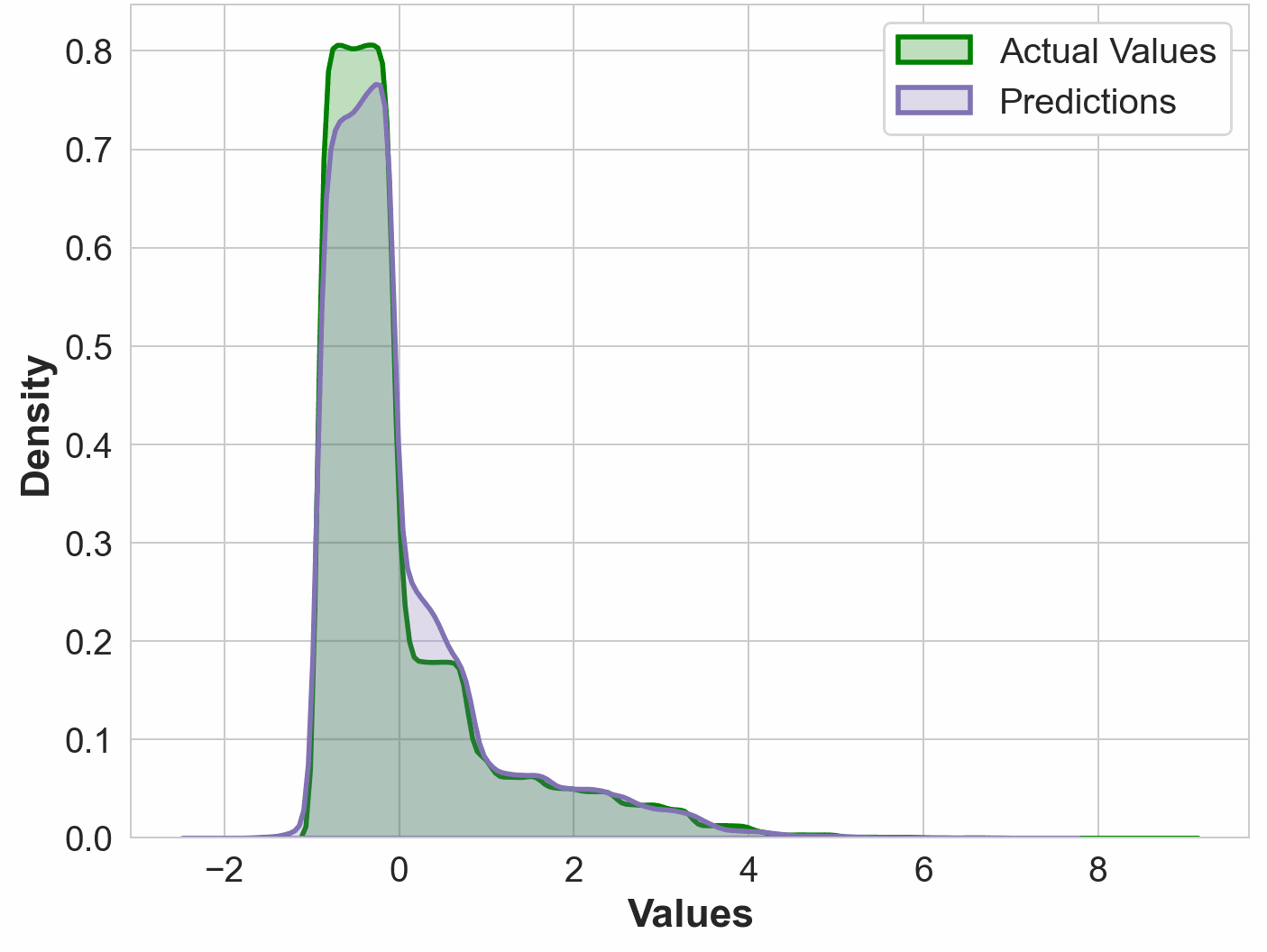}%
    \vspace{-0.5em}%
    \caption{Transformer model, \textit{compute\_time}}%
    \label{fig:eval-3-transformer-extrapolation-compute}  %
    \end{subfigure}%
    \hfill%
    \begin{subfigure}{0.49\linewidth}%
    \centering%
    \includegraphics[width=\linewidth]{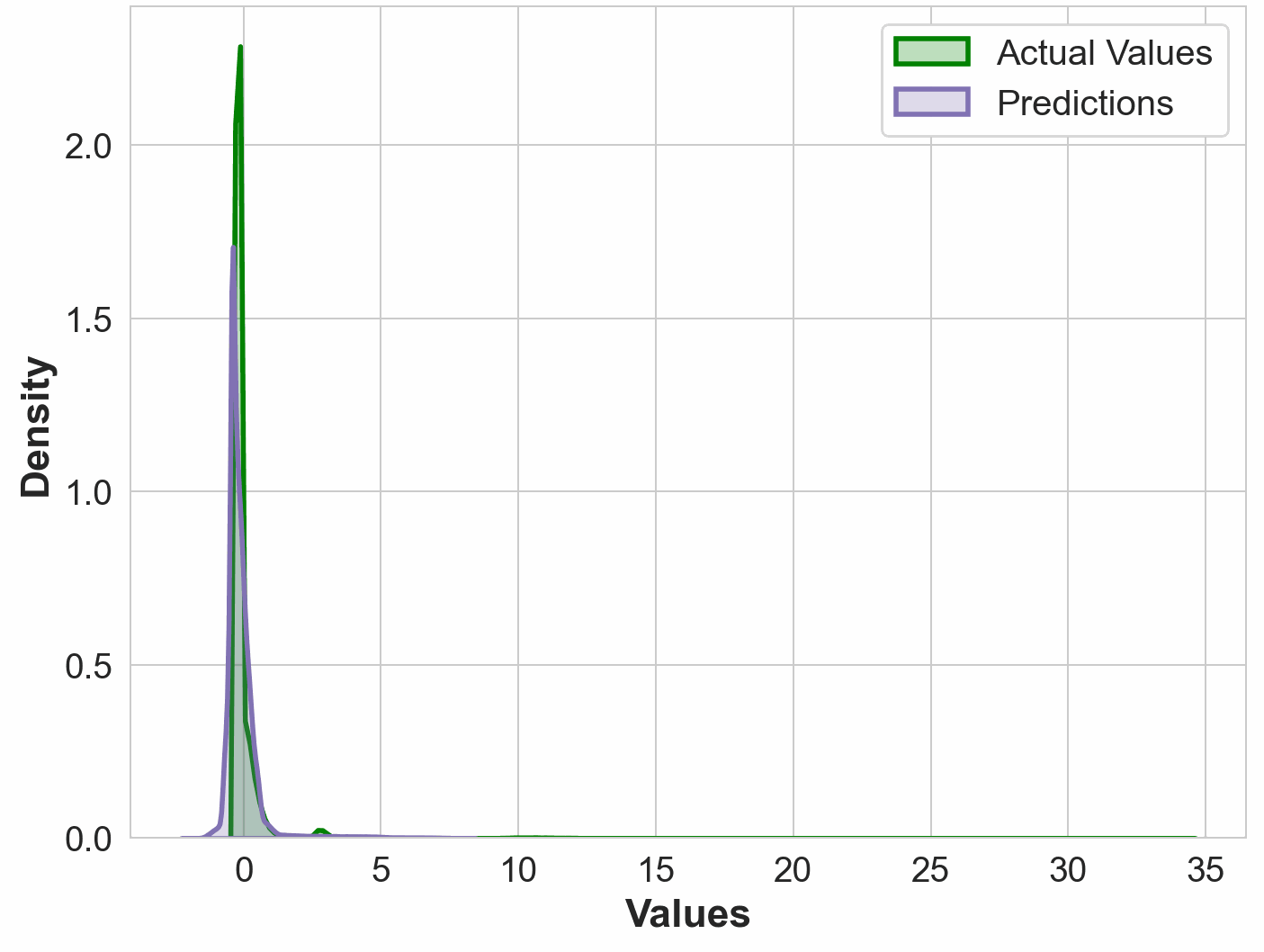}%
    \vspace{-0.5em}%
    \caption{Transformer model, \textit{input\_files\_transfer\_time}}%
    \label{fig:eval-3-transformer-extrapolation-input}%
    \end{subfigure}%
    \vspace{-1em}
    \caption{Predictions for extrapolation of trained models in the heterogeneous jobs setup.}%
    \label{fig:eval-3}%
\end{figure}

We present the prediction results 
in \autoref{tab:eval} and \autoref{fig:eval-3}.
We omit the KDE plots for the BiGRU model since they show the same features as the BiLSTM.
All models achieve an R-squared value close to one for \textit{compute\_time}, indicating near-optimal predictions. 
The KDE plots reveal smaller errors in matching the exact distribution but show a good fit overall despite the same lack of platform context as in the previous example. 
However, the R-squared value for \textit{input\_files\_transfer\_time} is negative across all three models, indicating that the models cannot predict the observable effectively. 
Looking at the KDE plots, we can see that the models accurately predict the spikes in the distribution. 
However, all models fail to predict the long tail of large values, deteriorating the overall prediction quality.

Overall, all models achieve similar results, with the Transformer model achieving slightly less accurate predictions. 
While \textit{compute\_time} can be predicted accurately, the behavior of \textit{input\_files\_transfer\_time} cannot be predicted accurately. 
This behavior may be influenced by complex interactions within the platform, such as the existence of various routes from the node storing the dataset to the worker node that processes the job. Since information about the platform's configuration is not explicitly included in the model inputs, it cannot be utilized to make predictions.
For the \textit{compute\_time}, however, the presence of multiple workloads, including jobs with different compute requirements and subsequently different execution times also depending on the worker they have been executed on, presents implicit platform information providing additional context about the platform.
For the \textit{transfer\_time}, however, the effect of the platform on the job execution is more convoluted and complex than what can be captured implicitly in the training data.
Nonetheless, the benefit of the surrogate in terms of execution times becomes more pronounced with a respective speed-up of multiple orders of magnitude ($\mathcal{O}(100\,\mathrm{s}) \to \mathcal{O}(100\,\mathrm{ms})$).
\begin{table}
    \centering%
    \caption{R-squared metric for different models and evaluation scenarios.}%
    \begin{adjustbox}{width=\linewidth}%
    \begin{tabular}{l|rrr|rrr}
    & \multicolumn{3}{c}{Homogeneous Jobs} & \multicolumn{3}{c}{Heterogeneous Jobs} \\
    Observable & BiGRU & BiLSTM & Transformer & BiGRU & BiLSTM & Transformer\\ \toprule
    \textit{compute\_time} & -1.20 & -1.08 & -0.14 & 0.99 & 0.99 & 0.96 \\
    \textit{input\_files\_transfer\_time} & 0.06 & 0.18 & 0.07 & -0.27 & -0.14 & -0.46 \\ \bottomrule
    \end{tabular}%
    \end{adjustbox}%
    \label{tab:eval}%
\end{table}%
\section{Conclusion}
\label{sec:conclusion}

We propose using surrogate modeling to evaluate the throughput of different infrastructure designs. 
We train three model architectures using simulator data to predict different job observables.
From our evaluation results, the architecture choice does not significantly influence the accuracy of the predictions at the current stage of development.
All three architectures decrease the execution times by orders of magnitude compared to DCSim.

At the current stage of the models inaccuracies are observed.
We suspect a lack of input information given to the models as the predominant source.
While the models are able to predict the compute times of our heterogeneous jobs scenario, where some implicit information about the infrastructure can be extracted by the model, they fail to predict the transfer time of input files due to not being aware of the data infrastructure setup that is more complex. 

Future work can build on these results by incorporating platform information into the training data to improve the predictions.
This will become essential to ensure that the model performs accurately on arbitrary workload mixes.
Moreover, training on real-world data instead of simulator data could enhance the capabilities and applicability of the models, also providing valuable data in regimes where the simulation, due to its scaling behavior, is not able to feasibly produce large amounts of training data.

\vspace{-.5em}
\section*{Acknowledgments}
This work was supported by the pilot program Core Informatics of the Helmholtz Association (HGF) and by the German Federal Ministry of Education and Research (BMBF) project FIDIUM 05H21VKRC2.
\vspace{-.5em}

\bibliography{bibliography} 
%
%
\end{document}